\documentclass[prb,twocolumn,superscriptaddress,showkeys]{revtex4-1}

\usepackage{graphicx}
\usepackage[colorlinks=true,linkcolor=blue,urlcolor=blue,citecolor=blue]{hyperref}

\begin{document}

\title{Direct characterization of plasmonic slot waveguides and nanocouplers}

\author{Andrei Andryieuski}
\email{andra@fotonik.dtu.dk}
\affiliation{DTU Fotonik, Technical University of Denmark, Oersteds pl. 343, DK-2800 Kongens Lyngby, Denmark}

\author{Vladimir A. Zenin}
\email{zenin@iti.sdu.dk}
\affiliation{Centre for Nano Optics, University of Southern Denmark, Campusvej 55, DK-5230 Odense M, Denmark}

\author{Radu Malureanu}
\affiliation{DTU Fotonik, Technical University of Denmark, Oersteds pl. 343, DK-2800 Kongens Lyngby, Denmark}

\author{Valentyn~S.~Volkov}
\affiliation{Centre for Nano Optics, University of Southern Denmark, Campusvej 55, DK-5230 Odense M, Denmark}

\author{Sergey I. Bozhevolnyi}
\affiliation{Centre for Nano Optics, University of Southern Denmark, Campusvej 55, DK-5230 Odense M, Denmark}

\author{Andrei V. Lavrinenko}
\affiliation{DTU Fotonik, Technical University of Denmark, Oersteds pl. 343, DK-2800 Kongens Lyngby, Denmark}

\date{\today}


\begin{abstract}
We demonstrate the use of amplitude- and phase-resolved near-field mapping for direct characterization of plasmonic slot waveguide mode propagation and excitation with nanocouplers in the telecom wavelength range. We measure mode's propagation length, effective index and field distribution and directly evaluate the relative coupling efficiencies for various couplers configurations. We report 26- and 15-fold improvements in the coupling efficiency with two serially connected dipole and modified bow-tie antennas, respectively, as compared to that of the short-circuited waveguide termination.
[This document is the unedited Author's version of a Submitted Work that was subsequently accepted for publication in \textit{Nano Letters}, \copyright American Chemical Society after peer review. To access the final edited and published work see \url{http://dx.doi.org/10.1021/nl501207u}.]
\end{abstract}

\maketitle

\begin{figure}
\centering\includegraphics[width=\linewidth]{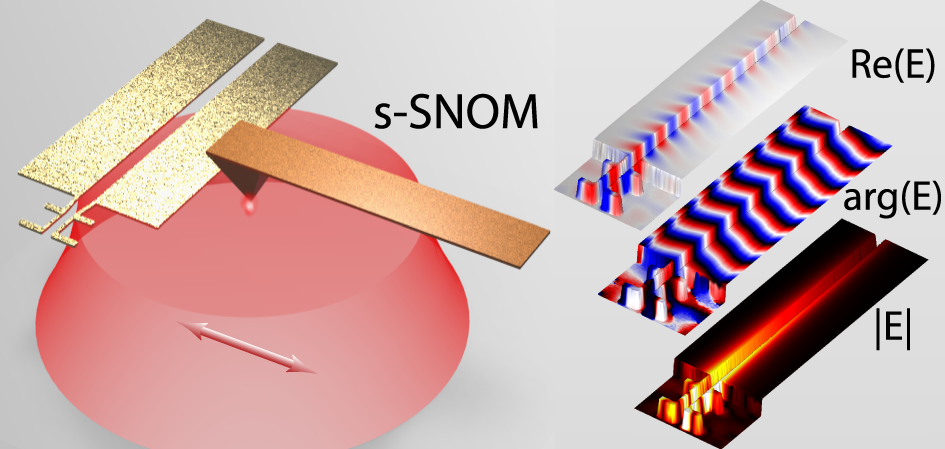}
\end{figure}

Great advantages offered by plasmonics to optical waveguiding are extreme subwavelength localization of guided modes close to the metal interface \cite{ref1} together with electrical tunability of electromagnetic waves via intrinsic metallic contacts \cite{ref2}. Plasmonic waveguides are therefore considered as a future generation of optical interconnects in integrated circuits for datacom technologies \cite{ref3}. Inevitably, with the appearance of nanoscale waveguides, a new challenge has emerged: how to effectively couple the diffraction-limited optical waves into deep-subwavelength plasmonic waveguides. Various approaches have been utilized ranging from lenses to grating couplers \cite{ref4}. However, the most compact solution is, an antenna based nanocoupler.

Antenna is a common tool to capture free-space propagating radio-waves with more than a century-long history \cite{ref5}. Employment of metal-based antennas in photonics started only in the last two decades owing to the progress in high-resolution  nanofabrication techniques \cite{ref6,ref7}. Usage of plasmonic antennas \cite{ref8} to couple light to plasmonic waveguides has been suggested theoretically \cite{ref9, ref10, ref11, ref12, ref13, ref14,ref15} and then confirmed experimentally with cross-polarization microscopy measurements in the near-infrared \cite{ref16} and with near-field microscopy in optical \cite{ref17}, telecom \cite{ref18} and mid-infrared \cite{ref19} ranges. Nevertheless, the amplitude- and phase-resolved measurements of the antenna-excited slot plasmons in the telecom range (with the free-space wavelength around 1.55 $\mu$m) have not been reported so far. It should be emphasized that the usage of {\it phase}-resolved near-field mapping is indispensable for {\it direct} characterization of the mode effective index as well as for revealing the {\it symmetry} of excited plasmonic modes \cite{ref19}.

In this Letter we report, for the first time to our knowledge, the amplitude- and phase-resolved near-field characterization of plasmonic slot waveguides and antenna based nanocouplers in the telecom wavelength range. Illumination with a wide laser beam excites both slot plasmons confined within a dielectric gap in a metal film and surface plasmon polaritons (SPP) propagating along the metal film interface perpendicular to the slot, and the resulting near-field interference pattern is mapped with a scattering-type scanning near-field optical microscope (s-SNOM). The observed interference pattern undergoes then the special filtration procedure in order to extract {\it individual} characteristics of the slot mode, including the effective index and propagation length, and its relative excitation efficiency, which is determined as a ratio between the slot mode intensity at the waveguide input and the average SPP intensity. Experimental characterization of two serial dipole and modified bow-tie antennas as well as the short-circuited waveguide termination in the absence of any coupling device is related to modelling of these configurations, including calculations of propagating mode fields and effective areas. 

\begin{figure*}
\centering\includegraphics{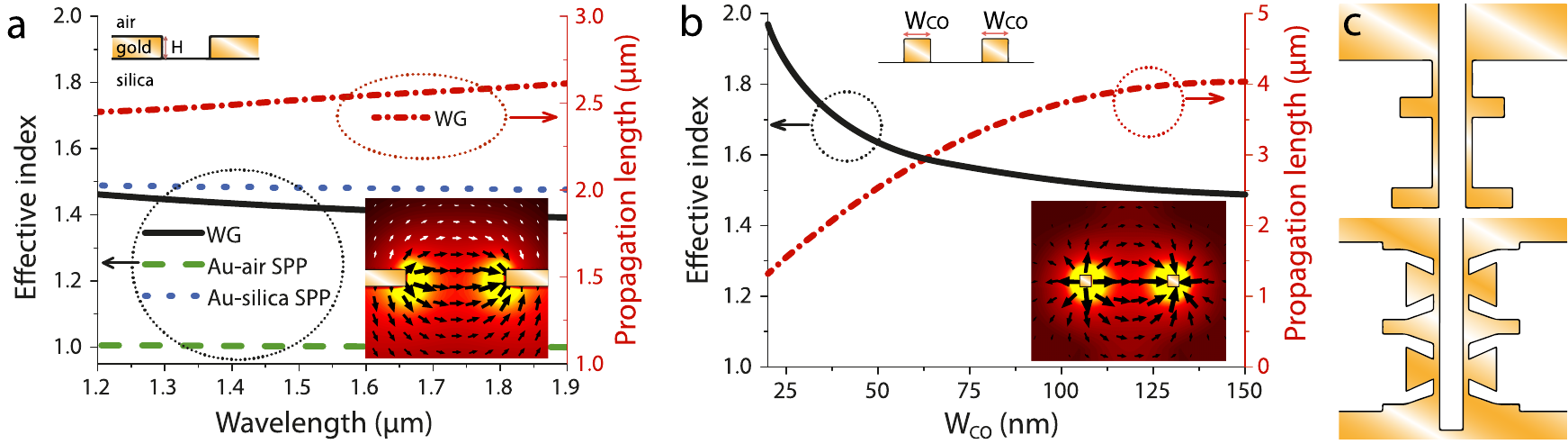}
\caption{(a) Effective indices of the slot waveguide mode (black solid line), gold-air (green dashed) and gold-silica (blue dotted) SPPs together with the propagation length of the slot mode (red dash-dotted). (b) Effective index (black) and propagation length (red dash-dotted) of the transmission line waveguide mode depending on the wire width $W_{\rm CO}$. Two serially connected (c)~dipole and (d) modified bow-tie antennas. Insets in (a) and (b) show cross-sections of the correspondent waveguides geometry and total electric field magnitude distributions of the plasmonic modes.}
\label{fig1}
\end{figure*}

We use a plasmonic slot waveguide \cite{ref20} (also known as a gap or channel waveguide), representing a rectangular slot of width $W_{\rm WG}$ = 300 nm carved in a gold film of thickness $H$ = 50 nm. Such a waveguide features both reasonably good mode confinement and propagation length. In a symmetric dielectric environment, the propagation length can reach few tens of micrometers at telecom wavelengths. We, however, select an asymmetric configuration (Figure~\ref{fig1}a, inset) that allows one to directly map plasmonic mode-field distributions with a sharp probe tip of the s-SNOM. The drawback of such configuration is the energy leakage from the slot mode to the slow SPPs on the silica-gold interface (the effective index of the latter is larger than that of the former). The leakage losses add up to the Ohmic losses in the plasmonic waveguide, resulting in the propagation length of ~2.5 $\mu$m (Figure~\ref{fig1}a). However, simulations indicate that it is sufficient to fill the slot space and add an additional 50 nm thick layer of silica on top in order to eliminate the leakage losses and increase the propagation length up to $\sim$7 $\mu$m. For comparison, the propagation lengths of the air-gold and silica-gold SPPs are 230 $\mu$m and 63 $\mu$m, respectively, at the wavelength 1.55 $\mu$m.

In our previous work \cite{ref9} we showed that the serial connection of the antennas provides no benefits for the slot excitation with the tightly focused Gaussian beam, whereas it is an opposite case for the incident plane wave or wide Gaussian beam used in our experimental setup. It is, therefore, important to ensure the optimal connection between the nanoantennas to deliver maximum of the incident energy to the waveguide. The effective index and propagation length of the double-wire transmission line depends on the wire width $W_{\rm CO}$ (Figure~\ref{fig1}b). Smaller losses correspond to wider wires, but wide wires connected to the antenna would prohibit efficient plasmons excitation. It is therefore better to increase the width of the connecting wires in the regions between the antennas, while keeping them narrow otherwise. In order to check the advantage of wider wires we compare thin-wire-connected dipole antennas (Figure~\ref{fig1}c) with modified bow-tie antennas that feature widened wire sections between them (Figure~\ref{fig1}d). To prevent the wide regions from out-of-phase plasmons excitation, we tune the length of the wide sections out of the resonance.

\begin{figure}[hb]
\centering\includegraphics{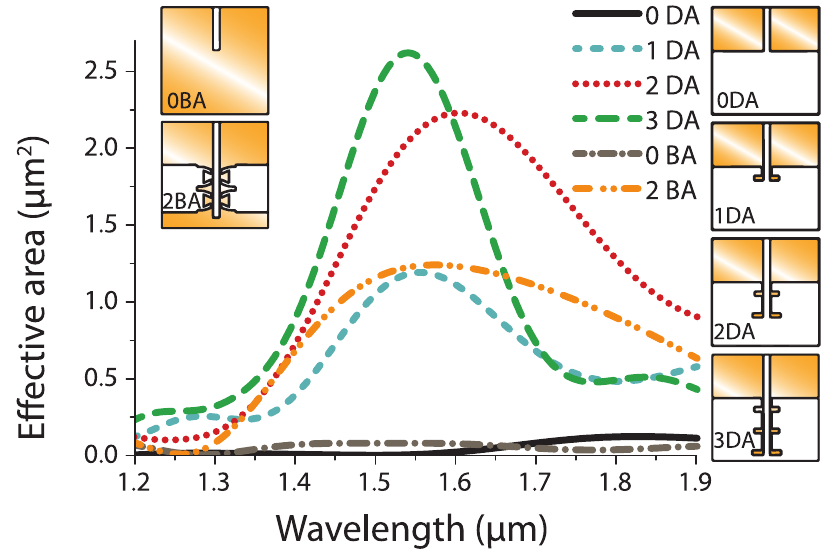}
\caption{Effective area of open circuit waveguide (0DA, black solid), one (1DA, cyan short-dashed), two (2DA, red dotted) and three (3DA, green long-dashed) serial dipoles, short-circuited waveguide (0BA, grey dash-dotted) and two serial modified bow-tie (2BA, orange dash-double-dotted) antennas. It is seen that the effective area $A_{\rm eff}$ increases with the number of antennas. The values  $A_{\rm eff}$ at the wavelength of 1.55~$\mu$m are collected in Table~\ref{table1}.}
\label{fig2}
\end{figure}

Numerical simulation and optimization of the nanocouplers for operation at the wavelength of 1.55 $\mu$m (see Supporting information for details) was carried out in CST Microwave Studio \cite{ref21}. For the plane wave excitation (normal to the substrate in our case), the figure-of-merit of the antenna, which characterizes its coupling efficiency and which is used as an objective function during optimization, is its effective area defined as the ratio of the power delivered to a waveguide mode to the incident power flux $A_{\rm eff} = P_{\rm WG}/S_{\rm inc}$. The effective area of one (1DA), two (2DA) and three (3DA) dipole antenna couplers was calculated and found being considerably and progressively larger than that of the open circuit waveguide termination (0DA) (Figure~\ref{fig2}, Table~\ref{table1}). At the same time, the operation full-width-at-half-maximum (FWHM) bandwidth of the 3DA coupler is smaller than that of the 2DA one. This fact agrees with the usual trade-off between antenna efficiency and bandwidth. The advantage of larger plasmon propagation length in wider connecting wires does not compensate for worse plasmon excitation due to interaction of metallic bars with the antenna elements and their detuning as well as counter-phase slot plasmons excitation in the modified bow-tie (2BA) antennas, resulting in $\sim$1.7 times smaller effective area, but $\sim$1.5 times larger bandwidth with respect to that of 2DA. The short-circuited waveguide termination (0BA) was found to be significantly more (by one order of magnitude) efficient for slot plasmon excitation that the 0DA configuration. Such a large difference in the coupling efficiencies can be explained by the difference in the corresponding excitation channels: in 0BA case, the waveguide mode is excited by directional light scattering from a metallic edge extending over the whole waveguide width, whereas, in 0DA case, it is only excited by scattering off metallic corners, with most of scattered light propagating away from the waveguide entrance. Bearing this in mind, we selected the 0BA configuration for the experimental fabrication and characterization together with the 2BA and 2DA cases that promise, respectively, 15- and 26-fold improvements in the effective area $A_{\rm eff}$ at the wavelength 1.55 $\mu$m (Table~\ref{table1}).

\begin{table}
\center
\caption{\label{table1} Bandwidth of the nanocouplers and their effective area at the optimization wavelength 1.55 $\mu$m .}
\begin{tabular}{ccc}
  \hline
  Design & $A_{\rm eff} (\mu{\rm m}^2)$ & Bandwidth ($\mu$m) \\
	\hline
	0DA & 0.0070 & \\
	1DA & 1.19 & 0.30\\
	2DA & 2.09 & 0.39\\
	3DA & 2.61 & 0.22\\
	0BA & 0.081 & 0.38\\
	2BA & 1.23 & 0.62\\
	\hline
 	 \end{tabular}
\end{table}

The nanocouplers were fabricated using standard electron-beam (e-beam) lithography followed by metal e-beam evaporation and lift-off in acetone (see Supporting information for details).

\begin{figure}[htb]
\centering\includegraphics{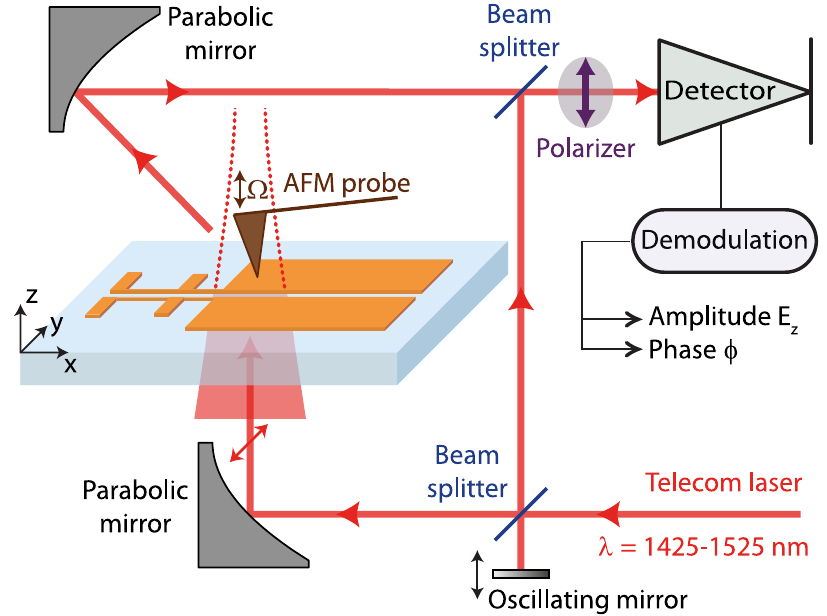}
\caption{Schematic of amplitude and phase near-field mapping with the transmission s-SNOM set-up. The sample is illuminated from below with a defocused laser beam (FWHM $\sim$ 12 $\mu$m) polarized parallel to the dipole antenna. The AFM metal-covered silicon tip scatters the near field (predominantly its vertical component), and the scattered radiation (being collected by the top parabolic mirror)  is then mixed with the reference beam and interferometrically detected, yielding amplitude and phase near-field distributions by scanning the sample.}
\label{fig3}
\end{figure}

Phase- and amplitude-resolved near-field characterization of the plasmonic antenna nanocouplers was carried out using the s-SNOM configuration, based on an atomic force microscope (AFM) with cantilevered tips being employed as near-field probes (NeaSNOM from Neaspec GmbH) (Figure~\ref{fig3}). In our experiments, we used standard commercial Si tips covered with platinum (Arrow NCPt, NanoWorld). The AFM was operated in the tapping mode, with the tip oscillating at the mechanical resonance frequency $\Omega \approx 250$~ kHz with the amplitude of $\sim$50~nm. Near-field and topography mapping was performed by moving the sample across the aligned configuration of the oscillating tip and the illumination system. Therefore, in order to continuously excite the antenna, we illuminated the structures from below (transmission-mode s-SNOM \cite{ref22}) with a defocused polarized laser beam (the estimated FWHM illumination spot was $\sim$12 $\mu$m). The light, scattered by the tip, was collected by a top parabolic mirror and directed towards a detector, where it was spatially overlapped with an interfering reference beam, yielding both the amplitude and phase of the scattered light via pseudo-heterodyne detection \cite{ref23}. Background contributions were suppressed by demodulating the detector signal at a high-order harmonic frequency $n\Omega$ (in our case $n = 2$), providing background-free near-field amplitude and phase images. It should be pointed out that, in most s-SNOM experiments, the illumination is done in the reflection mode (side-illumination scheme), where the incident light is focused on the tip with the same parabolic mirror that collects scattered light, a configuration that creates many problems for obtaining clear near-field images due to strong tip-sample coupling \cite{ref24} and phase-retardation effects \cite{ref25}. However, in our transmission-mode configuration, the sample was illuminated from below with an {\it in-plane} direction of polarization, allowing us to achieve uniform illumination and efficient excitation of the plasmonic antenna while avoiding the {\it direct} tip excitation \cite{ref22,ref25,ref26}. Due to a dominating dipole moment of the tip along its axis (i.e., along the $z$-axis), the recorded s-SNOM images represent mostly a distribution of the amplitude and phase of the $z$-component of the electric field, $E_z$ \cite{ref25}. In order to enhance this selectivity, a polarizer in front of the detector was set to select the {\it out-of-plane} $z$-polarization of light scattered by the tip. Finally, the recorded data were treated with free (scanning probe microscopy) software Gwyddion \cite{ref27}.

\begin{figure*}
\centering\includegraphics[width=0.75\textwidth]{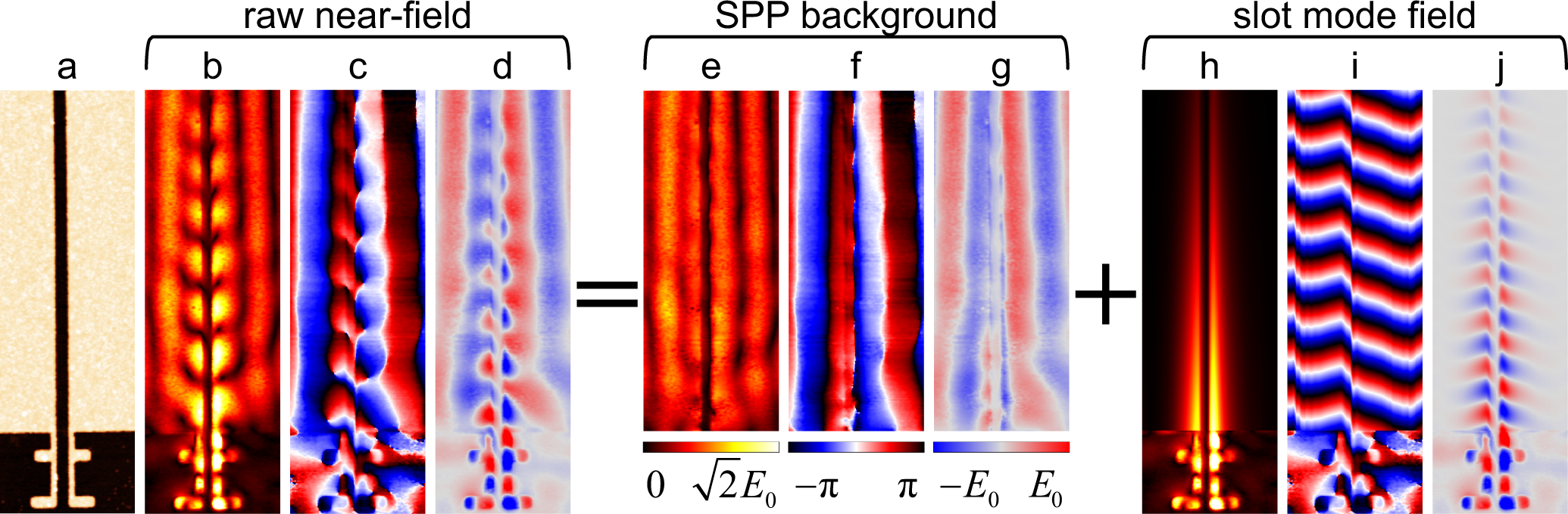}
\caption{Pseudocolor s-SNOM images, representing (a) topography, (b) amplitude, (c) phase, and (d) real part of the raw optical near-field distribution. Amplitude, phase, and real part of the decomposed contributions of (e-g) SPP background and (h-j) slot mode fields.}
\label{fig4}
\end{figure*}

\begin{figure*}
\centering\includegraphics[width=0.75\textwidth]{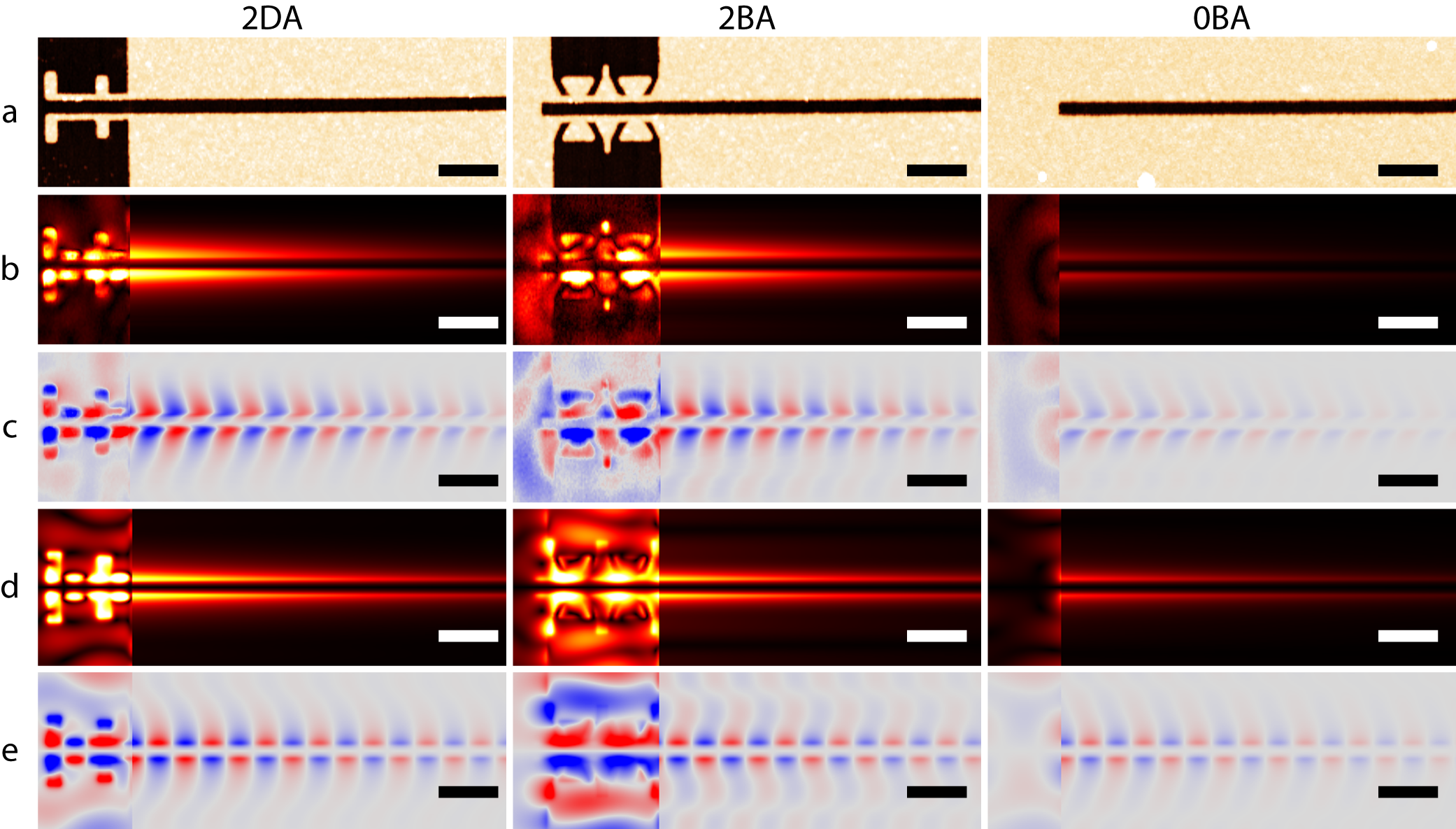}
\caption{Pseudocolor s-SNOM images, representing (a) topography, filtered (b) amplitude, and (c) real part of measured near-field data, and filtered (d) amplitude and (e) real part of simulated field for all three types of nanocouplers: 2DA (left), 2BA (middle) and 0BA (right). Electric field of the slot mode is normalized to the average amplitude of the background SPP. The scale bar is equal to 1 $\mu$m.}
\label{fig5}
\end{figure*}

The recorded optical amplitude and phase images (Figure~\ref{fig4}b,c) exhibit a complex interference pattern, produced mainly by the slot waveguide mode and SPPs. The latter are excited with the incident wave being diffracted on the slot, and propagate away from and perpendicular to the slot waveguide. Due to the large (defocused) excitation laser spot, adjusted with the AFM tip during the scan, the SPP amplitude and phase do not significantly change along the waveguide, as opposed to those of the slot mode. One can therefore decompose the recorded raw near-field data (Figure~\ref{fig4}b-d) in the half-space containing the waveguide into the SPP background (Figure~\ref{fig4}e-g) and slot mode (Figure~\ref{fig4}h-j) fields by fitting the data along the waveguide as a sum of propagating mode and a constant background (see Supporting information for details). Thus filtered images reveal the propagating slot mode with the decreasing amplitude (Figure~\ref{fig4}h) and linearly advancing phase (Figure~\ref{fig4}i) of $E_z$ field. As expected from the mode field distribution (Figure~\ref{fig1}a, inset), $E_z$ field magnitude is zero in the middle and opposite in sign on both sides of the waveguide (Figure~\ref{fig4}j).

The filtering procedure was applied to all measurements (see Supporting information for more details) made in the telecom range (1.425-1.525 $\mu$m) and to all three types of antenna nanocouplers: 2DA, 2BA, and 0BA (Figure~\ref{fig5}b-c). Numerically simulated $E_z$ field distribution at the height of 50 nm above the structure (the average position of the s-SNOM tip) was filtered with the same procedure (Figure~\ref{fig5}d-e), signifying very reasonable agreement with the experimental results. We should mention that phase images not only allow observing slot plasmon phase evolution, but also reveal information on the impedance matching of the antennas to the waveguide.

\begin{figure}[htb]
\centering\includegraphics{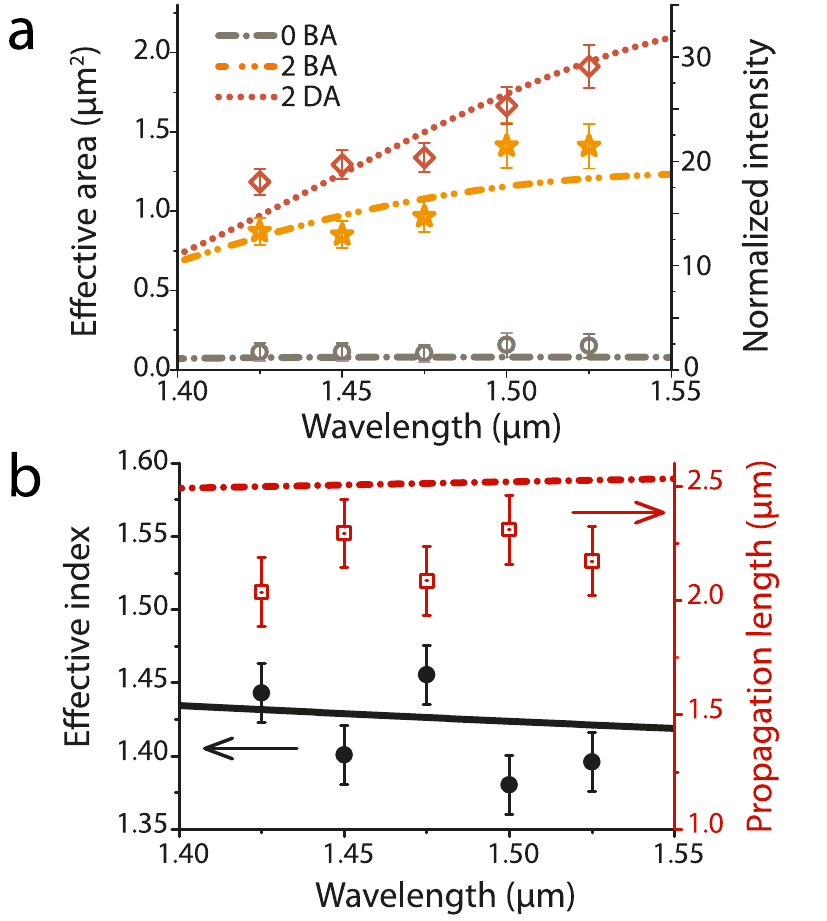}
\caption{(a) Wavelength dependence of the coupling efficiency of different type of antennas: 0BA (grey, circles), 2BA (orange, stars), and 2DA (red, rhombs), represented as effective area for numerical calculations (lines) and as normalized intensity for experimental results (points). (b) Wavelength dependence of the experimentally measured effective mode index (black circles) and propagation length (red squares), compared with numerical simulations (lines).}
\label{fig6}
\end{figure}

The filtered images of the slot mode allowed us to estimate the effective mode index, propagation length and intensity of the slot mode at the waveguide entrance (Figure~\ref{fig6}) normalized to the average intensity of the SPP background, which does not significantly depend on the wavelength in the range of 1.425-1.525 $\mu$m. Thus normalized slot mode intensity is expected to be proportional to the nanocoupler effective area $I_{{\rm slot\ at\ }x = 0} \sim P_{\rm WG} \sim A_{\rm eff}$.The validation of such approach is supported by the comparison of simulation results (lines) with experimental results (points), shown in Figure~\ref{fig6}a. The effective indices determined from the experimental (filtered) phase images (Figure~\ref{fig5}c) were found in good correspondence with the calculated ones (Figure~\ref{fig6}b), while the experimentally obtained propagation lengths turned out being slightly smaller than the calculated ones (Figure~\ref{fig6}b), most probably due to the fabrication imperfections. One may notice small oscillations in the experimentally measured values, which, we believe, represent a systematic error due to degradation or a drift in our setup, since a sequence of measurements for each antenna was 1475-1425-1525-1500-1450 nm, which correlates well with the error in propagation length and effective index (Fig.~\ref{fig6}b).

In conclusion, we have demonstrated the use of amplitude- and phase-resolved near-field mapping for complete characterization of the complex plasmonic waveguide configuration including antenna-based nanocouplers and slot waveguides. The s-SNOM characterization allowed us not only to make relative comparison of the efficiencies (in terms of the effective area) of different couplers, but also to measure the effective index and propagation length of the slot waveguide mode. All experimental data were found being in a very good correspondence with the numerical simulations. We have also confirmed that the serially connected dipole antennas represent the most efficient (for the excitation with a wide light beam) and simple design of nanocouplers. We would therefore anticipate that the serial antennas nanocouplers will become efficient optical interfaces between macroscopic light sources and nanoscale waveguides. We also believe that the s-SNOM-based characterization procedure described here will become a standard robust technique for the plasmonic waveguide characterization due to its high resolution, reliable measurements and efficient data filtration procedure.

\textbf{Acknowledgments}

A. A. acknowledges financial support from the Danish Council for Technical and Production Sciences through the GraTer project (Contract No. 0602-02135B). V. A. Z., V. S. V. and S. I. B. acknowledge financial support from the Danish Council for Independent Research (the FTP project ANAP, Contract No. 0602-01507B) and from the European Research Council, Grant No. 341054 (PLAQNAP). The authors also acknowledge J. Rosenkrantz de Lasson for a useful discussion on numerical simulations, G. Biagi, T. Holmgaard, J.-S. Bouillard and A. V. Zayats for discussions on near-field characterization and an anonymous reviewer for useful comments.

\textbf{Supporting Information Available}

Details on numerical simulation and optimization, the fabrication procedure, filtration procedure for experimental and simulated field distributions. This material is available free of charge via the Internet at \url{http://pubs.acs.org}.

\textbf{Author Contributions}
A. A. and V. Z. contributed equally.

\end{document}